\documentclass[twocolumn,pre,amsmath]{revtex4-1}
\usepackage[dvipdfmx]{graphicx}
\usepackage{graphicx}
\usepackage{epsfig,amsfonts,amsbsy,color}
\usepackage{amsmath,amssymb}
\usepackage{ascmac}
\usepackage{mathtools}
\usepackage{setspace}
\usepackage{algorithm}
\usepackage{algorithmic}

\usepackage{amsthm}
\theoremstyle{definition}

\newcommand{\var}{\mathrm{Var}}
\newcommand{\ex}{\mathbb{E}}

\begin{document}
\title{Kinetic uncertainty relation on first passage time for accumulated current}

\author{Ken Hiura}
\author{Shin-ichi Sasa}
\affiliation {
Department of Physics, Kyoto University, Kyoto 606-8502, Japan}

\date{\today}

\begin{abstract}
The kinetic uncertainty relation (KUR) is a trade-off relation between the precision of an observable and the mean dynamical activity in a fixed time interval for a time-homogeneous and continuous-time Markov chain. In this letter, we derive the KUR on the first passage time for the time-integrated current from the information inequality at stopping times. The relation shows that the precision of the first passage time is bounded from above by the mean number of jumps up to that time. We apply our result to simple systems and demonstrate that the activity constraint gives a tighter bound than the thermodynamic uncertainty relation in the regime far from equilibrium.
\end{abstract}

\maketitle

\section{Introduction}


The universal feature of dynamical fluctuations in systems far from equilibrium has been one of the main topics of nonequilibrium statistical physics over the past decades. A remarkable achievement in the field is the discovery of the fluctuation relations governing the fluctuation of the entropy production in generic stochastic systems, which are expressed as the equalities in contrast to the inequalities of the second law of thermodynamics. Recently, a novel inequality called the thermodynamic uncertainty relation (TUR) \cite{BS} has been studied; it provides an upper bound on the precision of the time-integrated current in terms of the mean entropy production. As well as the fluctuation relations, this type of inequality is universally valid for various stochastic systems under various situations \cite{GHPE, HGPRE17, DSJSM, PVdB, BCFG, KSP, TVVH, PS, DS}. See also \cite{HGNP}.


In a typical situation of these studies, we observe a stochastic system in a certain observation time interval and investigate the statistics of an observable at the fixed time. In this letter, we address a complementary problem, in which we exchange the roles of the observable and time, and study the statistics of the random time at which the observable first reaches a fixed threshold. Such random times are called first passage times (FPTs). The distributions of first passage times, or more generally stopping times, are extensively studied in various fields such as the theory of stochastic processes \cite{Redner,vanKampen}, reaction rate theory \cite{HTB}, biology \cite{IBZ, Ewens}, statistical estimations \cite{GMS}, and finance \cite{BBDG}. Moreover, in nonequilibrium physics, the universal natures of the first-passage-time statistics for thermodynamically relevant quantities were found recently, including the fluctuation relations at stopping times \cite{NRJ, MSMFPR}, the universality of the asymptotic behavior of the first-passage-time distributions \cite{SD, Ptaszynski, SMGKPFSRP}, and several tradeoff relations concerning the first passage time \cite{GH, Garrahan, FE}. In particular, the thermodynamic constraint on the precision of the FPT, which is analogous to the TUR on the precision of an accumulated current at a fixed time, may be useful in measuring the efficiency of the biological clocks beyond merely theoretical interests \cite{MCH} (see also \cite{BS, BSPRX, BSPRE}).


In this letter, we focus on the kinetic uncertainty relations (KURs) on the first passage times for time-integrated currents. Whereas the TUR gives a bound on the precision of an observable in terms of entropy production, the KUR \cite{TB} gives a bound in terms of the time-symmetric dynamical activity \cite{Maes}. Garrahan \cite{Garrahan} obtained a kinetic bound on the FPT for a stationary continuous-time Markov chain,
\begin{align} \label{eq:kur1}
 \frac{\ex [ \tau ]^2}{\var [ \tau ]} \leq n \ex [ \tau],
\end{align}
in the large threshold limit. Here $\ex [ \tau ]$ denotes the mean time at which the time-integrated current first reaches a threshold, $\var [ \tau ]$ the variance of $\tau$, and $n$ the mean number of jumps per time in the stationary state. This trade-off relation implies that the smaller the activity of the stochastic system, the larger the uncertainty of the time to reach the threshold. The inequality (\ref{eq:kur1}) was derived via the large deviation theory and verified only in the large threshold limit. The main purpose of the letter is to derive the KUR on the first passage time that is valid for any finite threshold and to simplify the derivation based on the technique recently developed in \cite{DS, TB}. The key ingredient of the derivation is the information inequality at stopping times.


The remainder of the paper is organized as follows. In Section \ref{sec:result}, we describe our setup and main result. In Section \ref{sec:app}, we apply our result to simple paradigmatic models and compare it with the TUR bound. In Section \ref{sec:derive}, we present a sketch of the proof of our result. In Section \ref{sec:concl}, we end with a summary and discussion. 

\section{Setup and Main Result}
\label{sec:result}

We consider a time-homogeneous and continuous-time Markov chain on a directed multigraph $G = (S, E)$. Here $S$ is a discrete state space and $E$ is the set of directed edges between two states. Let $k_e (x,y)$ be the transition rate from the state $x$ to $y$ via the edge $e \in E$ and $\lambda (x) \coloneqq \sum_{e} \sum_{y (\neq x)} k_e(x, y)$ the escape rate from $x$, where the summation is taken over edges starting from $x$. For a fixed time $t \geq 0$, $X_{[0,t]} = (X_s)_{s \in [0,t]}$ denotes a single trajectory of the system and is characterized by the discrete-time sequence $(x_0, t_0 = 0 ; x_1, t_1, e_1 ; \dots ; x_{N_t}, t_{N_t}, e_{N_t})$, which indicates that the total number of jumps for trajectory $X_{[0,t]}$ over $[0,t]$ is $N_t$, and the transition from $x_{i-1}$ to $x_{i} ( \neq x_{i-1})$ occurs via the edge $e_i$ at times $t_i$ for $i = 1, \dots, N_t$. We focus on a time-integrated current $J_t \coloneqq J(X_{[0,t]})$ defined as
\begin{align} \label{eq:current}
 J (X_{[0,t]}) = \sum_{i=1}^{N_t} g_{e_i} (x_{i-1}, x_i),
\end{align}
where $g_e(x,y)$ weights the contribution of the transition from $x$ to $y$ via the edge $e$. The class of observables of this form includes many important physical quantities. Here we address two significant examples, the number of jumps and the fluctuating entropy production. The total number of jumps via edge $f$ is obtained by taking $g_e = \delta_{e, f}$. This quantity measures how active the system is on edge $f$ and is called dynamical activity. Next, we consider two edges, $f$ connecting from $x$ to $y$, and $b$ connecting from $y$ to $x$, with $k_f(x,y) \neq 0$ and $k_b(y,x) \neq 0$. We assume that these edges are in contact with the same heat bath, and require that the entropy per the Boltzmann constant $k_{\mathrm{B}}$ produced in the heat bath during the transition $x \to y$ is given by $\ln ( k_{f} (x,y) / k_{b} (y,x))$. The fluctuating entropy production associated with the paired edges $(f,b)$ is then obtained by taking $g_e = \ln ( k_{f} (x,y) / k_{b} (y,x)) [\delta_{e, f} - \delta_{e,b}]$. The requirement we impose here is called the local detailed balance condition. However, in this letter, we do not impose any requirements on the weight function $g_e$ such as non-negativity, symmetry, and anti-symmetry.

The first passage time (FPT) $\tau$ for the time-integrated current $J_t$ is defined as $\tau \coloneqq \inf \{ t \geq 0 : J_t > J_{\mathrm{th}} \}$, where $J_{\mathrm{th}}$ denotes the threshold value. The FPT is obviously a stochastic variable and accompanies fluctuations. The first time for the system to reach the specific state $z$ can be represented in this form by taking $g_{e}(x,y) = \delta_{y,z}$ and $J_{\mathrm{th}} \in (0,1)$. Methods to analyze the statistics of the FPTs in the class are well established \cite{Redner, vanKampen}. Our concern here is the precision of the FPT quantified by the ratio of the squared mean FPT to the variance, $\ex [ \tau]^2 / \var [ \tau ]$. Throughout this paper, we use $\ex [ f ]$ to denote the expectation value of $f$ with respect to the underlying stochastic process and $\var [ f ] = \ex [ (f - \ex [ f ] )^2]$ the variance.

We suppose that the mean and variance of $\tau$ are finite. We find that the precision of the FPT is bounded from above by the mean dynamical activity, which is quantified by the mean number of jumps:
\begin{align} \label{eq:result}
 \frac{\ex_{x_0} [ \tau]^2}{\var_{x_0} [ \tau ]} \leq \ex_{x_0} [ N_{\tau} ],
\end{align}
where $N_{\tau}$ is the total number of jumps up to the first passage time $\tau$ and $\ex_{x_0} [ \cdot ]$ denotes the expectation value conditioned on the initial configuration $X_0 = x_0$. This activity bound (\ref{eq:result}) is the main result of this letter. The inequality (\ref{eq:result}) implies that the reduction in the number of jumps up to the time for the system to passage the threshold inevitably accompanies the worsening of the optimal precision of the FPT. We note that the inequality (\ref{eq:result}) is still valid if we replace the conditional expectation $\langle \cdot \rangle_{x_0}$ by the expectation value $\langle \cdot \rangle_{\rho}$ with respect to the arbitrary initial distribution $\rho$.

We make several remarks on our result. First, our result holds for any finite threshold $J_{\mathrm{th}}$ in contrast to the inequality (\ref{eq:kur1}) in \cite{Garrahan}. For a sufficiently large threshold $J_{\mathrm{th}}$ and ergodic Markov process, we expect that $\ex [ N_{\tau} ]$ nearly equals $n \ex [ \tau ]$ because in that situation the jump number per time is well approximated over a sufficiently large time interval by the stationary value $n$. Hence the inequality (\ref{eq:kur1}) is recovered in this asymptotic limit from our result. Although the precision of the FPT for the integrated current is also bounded from above by the mean entropy production, the thermodynamic bound of the form in \cite{GH} is guaranteed only in this asymptotic limit. We illustrate the violation of the TUR with examples in the next section. Second, the thermodynamic bound is tighter than the activity bound around equilibrium because in the regime close to equilibrium the mean entropy production tends to zero but the mean number of jumps remains finite. In contrast, even when the mean entropy production goes to infinity in the regime far from equilibrium, the mean activity up to the first passage time may be finite due to a nonequilibrium force driving the system to the threshold. In that case, the KUR provides a tighter bound than the TUR. Third, the KUR is applicable even if the Markov chain is not ergodic or does not satisfy the reversibility condition, i.e., $k(x,y) \neq 0$ iff $k(y,x) \neq 0$. Examples of stochastic processes violating the reversibility condition are models including absorbing states in the population dynamics. Our final remark is that the mean and variance of the FPT may diverge; in such circumstances, the KUR (\ref{eq:result}) may be violated. For instance, when the accumulated current has a positive drift, the FPT for a negative threshold takes infinite value with positive probability. However, the modified KUR still holds in the following form \cite{suppl},
\begin{align} \label{eq:result2}
 \frac{\widetilde{\ex}_{x_0} [ \tau ]^2}{\widetilde{\var}_{x_0} [ \tau ]} \leq \widetilde{\ex}_{x_0} [N_{\tau}].
\end{align}
Here $\widetilde{\ex} [ f ] \coloneqq \ex_{x_0} [f 1_{\{\tau < \infty \}} ]$ is the integration of $f$ over the restricted region $\{ \tau < \infty \}$ and $\widetilde{\var}_{x_0} [f] \coloneqq \widetilde{\ex} [ (f - \widetilde{\ex} [f])^2]$ the corresponding variance. We note that if the probability that the first passage time is infinite is positive, the modified probability distribution is unnormalized, $\widetilde{\ex} [ 1 ] = \mathrm{Prob}(\tau < \infty)< 1$.


\section{Applications}
\label{sec:app}

\begin{figure}
\centering
\includegraphics[width=7cm]{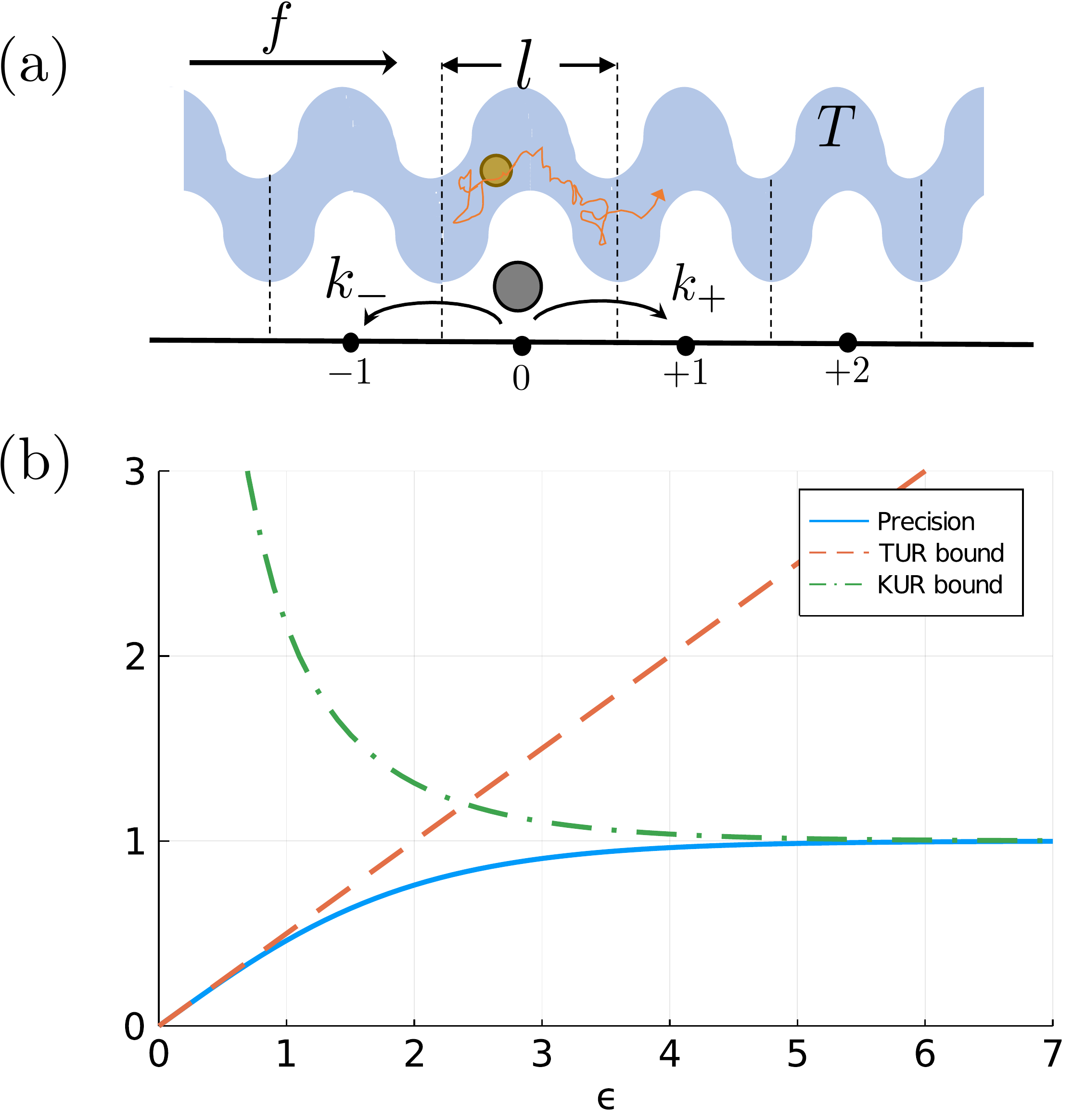}
\caption{(Color online) (a) Schematics of a colloid being driven by an external force $f$ in a periodic channel filled with water at temperature $T$ and 1D biased random walk describing the dynamics of the colloid. (b) Plots showing the precision $\ex [ \tau_x]^2/(x\var[\tau_x])$ (blue solid line), the TUR bound $\sigma \ex[\tau_x]/(2x)$ (orange dash line), and the KUR bound $\ex[N_{\tau_x}]/x$ (green dash-dot line) as functions of $\epsilon$. The dimensionless parameter $\epsilon = fl/k_{\mathrm{B}}T$ measures a distance from equilibrium in this model.}
\label{fig:srw}
\end{figure}

We examine the KUR (\ref{eq:result}) in two paradigmatic examples. The first example is the biased random walk $X_t$ on $\mathbb{Z}$ starting from $X_0 = 0$. The transition rates between neighboring sites are set to $k_{\pm} \coloneqq k(x, x \pm 1) = a e^{\pm \epsilon / 2}$ and other transitions do not occur. Here $a > 0$ and $\epsilon > 0$ are positive constants. Suppose that this system describes a colloid under an external driving force $f$ in a channel having a periodic structure of length $l$ and filled with water in equilibrium at temperature $T$ (Fig. \ref{fig:srw}(a)). According to the local detailed balance condition, $\epsilon = fl / k_{\mathrm{B}} T = \ln (k_{+} / k_{-})$ is the entropy per $k_{\mathrm{B}}$ produced in the water by the one forward jump. We consider the random time $\tau_{x} = \inf \{ t \geq 0 : X_t = x \}$ at which the colloid first reaches the site $x > 0$. The entropy production along the path $X_{[0,t]}$ is given by $\Sigma_t \coloneqq \epsilon X_t$ and therefore the stationary entropy production is $\sigma \coloneqq \ex [ \Sigma_t ] / t = 2 \epsilon \sinh(\epsilon / 2)$. We easily find that the precision of $\tau_x$ is given by $\ex [\tau_x ]^2 / \var [ \tau_x ] = x \tanh (\epsilon /2)$ and the TUR \cite{GH}, $\ex [\tau_x ]^2 / \var [ \tau_x ] \leq \sigma \ex [ \tau_x] / 2$, is directly verified for any thresholds \cite{GRJ}. In addition, the mean dynamical activity is given by $\ex [ N_{\tau_{x}} ] = x \coth (\epsilon / 2)$ and is in agreement with the KUR, $\ex [\tau_x ]^2 / \var [ \tau_x ] \leq \ex [ N_{\tau_{x}} ]$ \cite{suppl}. In Fig. \ref{fig:srw}(b), we see that while the TUR is tighter near the equilibrium $\epsilon \lesssim 1$, the KUR becomes relevant as the nonequilibrium driving force increases. We remark that the TUR may be violated in general for finite thresholds. As an example, we consider a random walk with a reflecting boundary condition at the origin, i.e., $k(0, -1) = 0$ and a precision of the FPT for the threshold $x=1$. The distribution of $\tau_1$ is the exponential distribution with the decay rate $k_{+}$ and therefore $\ex [\tau_1 ]^2 / \var [ \tau_1 ] = 1 = \ex [ N_{\tau_1} ]$ for any $\epsilon$, whereas the upper bound of the TUR, $\sigma \ex [ \tau_1] = \epsilon / 2$, becomes less than 1 for sufficiently small $\epsilon$.

\begin{figure}
\centering
\includegraphics[width=7cm]{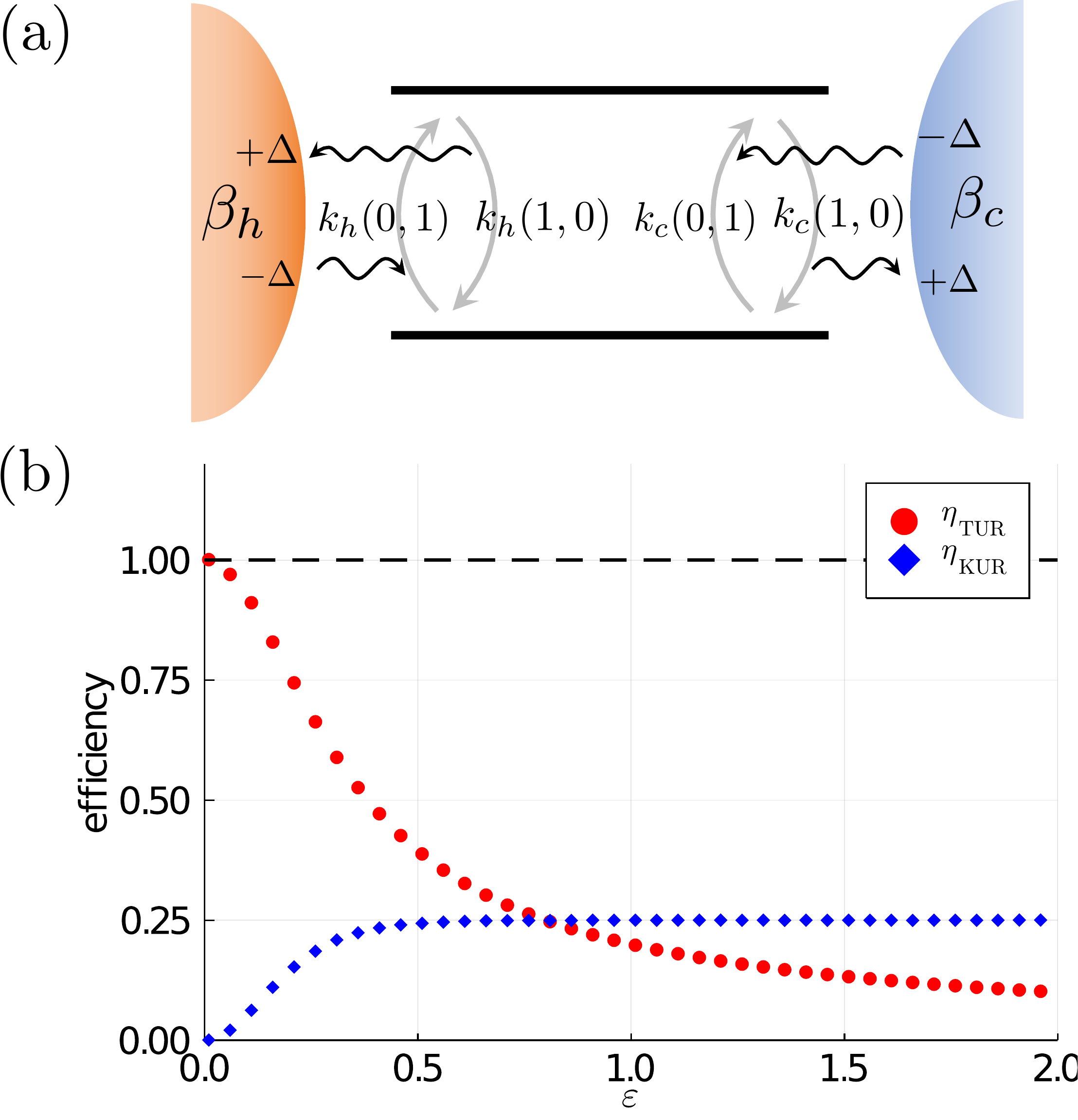}
\caption{(Color online) (a) Schematic of two-level system. (b) Efficiencies associated with the TUR (red circles) and KUR (blue diamonds) obtained from the Monte Carlo simulations for $10^7$ samples for various temperature differences. We set $X_0 = 0$, $k_c(0,1)=1$, $k_c(1,0) = e^{\beta_c \Delta}$, $k_h(0,1)=1$, $k_h(1,0)=e^{\beta_h \Delta}$, $\overline{\beta} \Delta = 10$, $m=1$ in our simulation.}
\label{fig:twolevel}
\end{figure}

The second example is a two-level system in contact with two heat baths at different temperatures (Fig. \ref{fig:twolevel}(a)). The lower (resp. higher) energy level is coded by $0$ (resp. $1$) and the energy gap is set to $\Delta > 0$. The transition rate from $x \in \{0,1\}$ to $y (\neq x)$ associated with the heat bath at the inverse temperature $\beta_{e}$ is given by $k_e(x,y) > 0$ for $e \in \{ h, c \}$. We assume that $\beta_h < \beta_c$ and define $\overline{\beta} \coloneqq (\beta_c + \beta_h)/2$. The local detailed balance condition imposes $k_e(1,0) / k_e(0,1) = e^{\beta_e \Delta}$ for each bath $e$. We observe the heat produced in the cold bath $e=c$ and measure the accumulated heat current per $\beta_c \Delta$ from the system into the cold bath $J_t \coloneqq \sum_{i=1}^{N_t} [ \delta_{x_{i-1},1} \delta_{x_i, 0} - \delta_{x_{i-1}, 0} \delta_{x_i, 1}] \delta_{e_i, c}$. Our interest is the variation in precision of the first passage time $\tau_m = \inf \{ t \geq 0 : J_t = m \}$ ($m \in \mathbb{Z}$) along with the temperature difference, which quantifies a distance from equilibrium in this model. To measure the degree by which temperatures differ, we introduce a dimensionless parameter $\epsilon \coloneqq (\beta_c - \beta_h)/ \overline{\beta} \in [0,2]$. We define the efficiencies associated with the TUR and KUR as
\begin{align}
 \eta_{\mathrm{TUR}} \coloneqq \frac{2}{\sigma} \frac{\ex [\tau_m]}{\var [ \tau_m ]}, \ \eta_{\mathrm{KUR}} \coloneqq \frac{1}{\ex [ N_{\tau_m} ]} \frac{\ex [\tau_m]^2}{\var [ \tau_m ]} \leq 1,
\end{align}
respectively. From the plot of the efficiencies obtained by Monte Carlo simulations (Fig. \ref{fig:twolevel}(b)), we see that although they are actually reversed as the temperature difference increases, the efficiency associated with the KUR is always far from optimal.

\section{Sketch of the derivation of (\ref{eq:result})}
\label{sec:derive}

We sketch a derivation of (\ref{eq:result}) via the information inequality at stopping times \cite{suppl}. Let us consider a family of stochastic processes with path probability distributions $P_{\theta}$ smoothly parametrized by a real parameter $\theta$. We use $p_{\theta}^t \coloneqq dP_{\theta}^t / dP_{0}^t$ to denote the likelihood ratio function up to time $t$ with respect to the reference process $\theta = 0$. A stopping time $\tau$ is defined informally as a $[0, \infty]$-valued random time that does not depend on the trajectory in the future $X_{(\tau,\infty)}$. Typical examples of stopping times are first passage times. Under several regularity conditions, the information inequality \cite{Magiera, FM} claims that for a general non-anticipating observable $A_t$ and a stopping time $\tau$ with $P_{\theta} \{ 0 < \tau < \infty \} = 1$, the following inequality holds;
\begin{align} \label{eq:cr}
 \frac{ ( \partial_{\theta} \ex_{\theta} [ A_{\tau} ])^2}{ \var_{\theta} [ A_{\tau} ]} \leq I_{\tau} (\theta).
\end{align}
Here $\ex_{\theta} [ \cdot ]$ denotes the expectation with respect to $P_{\theta}$ and $I_{\tau}(\theta) = \ex_{\theta} [ - \partial_{\theta}^2 \ln p_{\theta}^{\tau} ]$ the Fisher information at the stopping time $\tau$. We apply this inequality to the first passage time for a time-integrated current $J_t$ and a family of continuous-time Markov chains with the transition rates $k_{e,\theta} = (1+\theta) k_{e}$ starting from the initial condition $X_0 = x_0$. The transformation of the transition rates $k \mapsto (1+\theta)k$ corresponds to a rescaling of the time scale of the stochastic evolution. Therefore, the expectation of the FPT for the observable of the form (\ref{eq:current}) transforms as $\ex_{\theta,x_0}[ \tau ] = (1+\theta)^{-1} \ex_{x_0}[ \tau ]$. The log-likelihood ratio function is given by
\begin{align}
 \ln p_{\theta}^t = N_t \ln (1+\theta) - \theta \int_{0}^{t} \lambda (X_s) ds
\end{align}
from the Girsanov formula \cite{KL}. The Fisher information for this family is thus obtained as $I_{\tau}(\theta) = (1+\theta)^{-2} \ex_{x_0} [ N_{\tau} ]$. By substituting these results into Eq. (\ref{eq:cr}) and setting $\theta=0$, we have the KUR (\ref{eq:result}). It is straightforward to include initial distributions in the derivation.

We make a remark on the derivation of the thermodynamic uncertainty relation from the information inequality. In Ref. \cite{DS}, Dechant and Sasa derived the finite-time TUR on a time-integrated current from the information inequality. Specifically, they found a family of stationary continuous-time Markov chains $\{ P_{\theta} \}$ for which $\ex_{\theta}[J_t] = (1+\theta) \ex [J_t]$ and $I_t (\theta) \leq \ex [\Sigma_t] / 2$, where $\Sigma_t$ is the total entropy production over $[0,t]$. Although one may expect that the mean first passage time for $J_t$ has rescaling property $\ex_{\theta}[\tau] = (1+\theta)^{-1} \ex [\tau]$ for this family and the finite-threshold TUR on the FPT can be derived from (\ref{eq:cr}), it is not true. This is because the statistics of the first passage time depends on the transition probability. Although the perturbation considered in \cite{DS} corresponds to the time-rescaling of the single-time probability distribution and current, it does not have the same property at the level of the transition probability.

\section{Discussion}
\label{sec:concl}

We have derived the kinetic bound (\ref{eq:result}) on the first passage time for time-integrated current that is valid for any finite threshold. Contrary to TURs, KUR may be relevant for a system far from equilibrium. An interesting challenge is to apply our result to biological systems such as circadian clocks and molecular motors and measure the efficiency of these systems from the perspective of the precision of the first passage time.

Refs. \cite{GH, Garrahan} use the connection between the rate functions for current statistics and first-passage-time statistics to derive the same type of inequality. Nevertheless, our derivation is based on the idea given in \cite{DS, TB} that finite-time TUR and KUR are obtained from the information inequalities for virtually perturbed systems. This method significantly simplifies the derivation and extends the range of applicability.

Our result is widely applicable to generic continuous-time and time-homogeneous Markov chains including non-thermodynamic systems. Extending the KUR on the FPT to diffusion processes and quantum systems under continuous measurement \cite{Hasegawa} remains an open problem.

\section*{Acknowledgment}
The authors thank Andreas Dechant for valuable discussions. KH was supported by JSPS KAKENHI Grant Number 20J12143. SS was supported by JSPS KAKENHI Grant Numbers JP17H01148, JP19H05795, and JP20K20425.

\clearpage

\begin{widetext}

\appendix
\def\thesection{\Alph{section}}

\begin{center}
\textbf{Supplemental Material for \\``Kinetic uncertainty relation on first passage time for accumulated current''}

Ken Hiura and Shin-ichi Sasa

\textit{Department of Physics, Kyoto University, Kyoto 606-8502, Japan}
\end{center}

In Section \ref{ap:1}, we review for the readers' convenience information inequalities at stopping times. In Section \ref{ap:2}, we present a derivation of the modified KUR (\ref{eq:result2}) including Eq. (\ref{eq:result}) as a special case. In Section \ref{ap:3}, we describe how to calculate the mean dynamical activity up to the first passage time for random walks.

\section{Information inequalities at stopping times}
\label{ap:1}

We review the information inequalities at stopping times used in our derivation albeit without mathematical rigor. See \cite{Magiera, FM, KS} for details of the mathematical background and regularity conditions that are not explicitly mentioned below. We consider a family of stochastic processes over time interval $[0,\infty)$ with path probability distributions $P_{\theta}$ parametrized by a real parameter $\theta$. This situation is mathematically modeled by the probability space $(\Omega, \mathcal{F}, \{ \mathcal{F}_t \}_{t \geq 0}, P_{\theta})$. Here $\Omega$ denotes the space of stochastic trajectories over $[0,\infty)$, $\mathcal{F}$ the collection of all events that can occur in this system, $\mathcal{F}_t (\subset \mathcal{F})$ the collection of events that can occur up to fixed time $t$, and $P_{\theta} : \mathcal{F} \to [0,1]$ the map assigning events in $\mathcal{F}$ into probabilities. $\mathcal{F}$ and $\{ \mathcal{F}_t \}_{t \geq 0}$ are called the $\sigma$-algebra and filtration, respectively. We use $P_{\theta}^t$ to denote the probability distribution that assigns probabilities to events in $\mathcal{F}_t$. More formally, $P_{\theta}^t$ is defined as the restriction of $P_{\theta}$ to $\mathcal{F}_t$, i.e., $P_{\theta}^t \coloneqq P_{\theta} |_{\mathcal{F}_t}$. We assume that for any $\theta$, $\theta^{\prime}$ and $t \geq 0$, $P_{\theta}^t$ are mutually absolutely continuous with respect to $P_{\theta^{\prime}}^t$. Under this assumption, we define the likelihood ratio function $p_{\theta}^t \coloneqq dP_{\theta}^t / dP_0^t$ for each $t \geq 0$.

Let $\tau$ be a stopping time for this process. This means that for each $t \geq 0$, whether the value of $\tau$ is smaller than $t$ is determined by the information about the trajectory up to the time $t$, i.e., $\{ \tau \leq t \} \in \mathcal{F}_t$. We define the $\sigma$-algebra $\mathcal{F}_{\tau}$ generated by the stopping time $\tau$ as
\begin{align}
 A \in \mathcal{F}_{\tau} \Leftrightarrow (\forall t \geq 0) ( A \cap \{ \tau \leq t \} \in \mathcal{F}_t).
\end{align}
This means that $\mathcal{F}_{\tau}$ is the collection of events of which we can determine the happening before the stopping time $\tau$. For instance, whether the observable at the time $\tau$, $A_{\tau}$, is in a certain range $[a, a+da)$ is determined by the information on the trajectory up to the time $\tau$ and therefore $\{A_{\tau} \in [a, a+da) \} \in \mathcal{F}_{\tau}$. The probability distribution $P_{\theta}^{\tau}$ that assigns probabilities to events in $\mathcal{F}_{\tau}$ is defined as the restriction of $P_{\theta}$ to $\mathcal{F}_{\tau}$, i.e., $P_{\theta}^{\tau} \coloneqq P_{\theta}|_{\mathcal{F}_{\tau}}$. The expectation value of an observable $A_{\tau}$ at the stopping time is obtained via this distribution $P_{\theta}^{\tau}$: $\ex_{\theta} [ A_{\tau} ] = \int A_{\tau} dP_{\theta}^{\tau}$. 

Whether the random variable $p_{\theta}^{\tau} (\omega) \coloneqq p_{\theta}^{\tau(\omega)}(\omega)$ is related to the Radon-Nikodym derivative of $P_{\theta}^{\tau}$ with respect to $P_0^{\tau}$ is a non-trivial problem because these two objects are defined in different ways.
It is known \cite[Appendix B, Theorem B.1.1]{KS} that
\begin{align} \label{eq:sudakov}
 P_{\theta} ( E \cap \{ \tau < \infty \}) = \int_{E} p_{\theta}^{\tau} 1_{\{ \tau < \infty \}} dP_0
\end{align}
for any events $E \in \mathcal{F}_{\tau}$. By defining the modified probability distribution $\widetilde{P}_{\theta} [ E ] = P_{\theta} [ E \cap \{ \tau < \infty \} ]$ and the corresponding expectation $\widetilde{\ex}_{\theta}$, we obtain that
\begin{align} \label{eq:sudakov2}
 \widetilde{\ex}_{\theta} [ A_{\tau} ] = \widetilde{\ex}_0 [ A_{\tau} p_{\theta}^{\tau} ]
\end{align}
from (\ref{eq:sudakov}). We remark that $\widetilde{\ex}_{\theta} [1] = P_{\theta}(\tau < \infty)$ may be less than 1, i.e., the distribution $\widetilde{P}_{\theta}$ may be unnormalized. If $P_{\theta} (\tau < \infty) = 1$ for any $\theta$, we have that $P_{\theta}^{\tau}$ is mutually absolutely continuous with respect to $P_{\theta^{\prime}}^{\tau}$ and the Radon-Nikodym derivative is actually related to $p_{\theta}^{\tau}$ \cite[Theorem B.1.2]{KS}:
\begin{align}
 p_{\theta}^{\tau} = \frac{dP_{\theta}^{\tau}}{dP_0^{\tau}}.
\end{align}
In that case, $\widetilde{P}_{\theta}$ and $\widetilde{\ex}_{\theta}$ yield the normalized probability distribution and the usual expectation value. 

Based on these results, we explain the information inequalities at stopping times. A crucial assumption here is that $\widetilde{\ex}_{\theta} [1]$ is independent of $\theta$, i.e.,
\begin{align} \label{eq:indep}
 \partial_{\theta} \widetilde{\ex}_{\theta}[1] = 0.
\end{align}
This assumption means that the probabilities that the stopping time diverges have the same value over all parameters. Under this assumption and several regularity conditions, we obtain the inequality,
\begin{align}
 \partial_{\theta} \widetilde{\ex}_{\theta}(A_{\tau}) &= \widetilde{\ex}_0 [ p_{\theta}^{\tau} (\partial_{\theta} \ln p_{\theta}^{\tau}) A_{\tau} ]
 \notag \\
 &= \widetilde{\ex}_{\theta} [ (\partial_{\theta} \ln p_{\theta}^{\tau}) A_{\tau} ]
 \notag \\
 &= \widetilde{\ex}_{\theta} [ (\partial_{\theta} \ln p_{\theta}^{\tau}) (A_{\tau} - \widetilde{\ex}_{\theta} [ A_{\tau} ])]
 \notag \\
 & \leq \sqrt{\widetilde{\ex}_{\theta} [ (\partial_{\theta} \ln p_{\theta}^{\tau})^2]} \sqrt{\widetilde{\var}_{\theta}[A_{\tau}]},
\end{align} 
where we have used (\ref{eq:sudakov2}) and $\partial_{\theta} p_{\theta}^{\tau} = p_{\theta}^{\tau} \partial_{\theta} \ln p_{\theta}^{\tau}$ in the first line, (\ref{eq:sudakov2}) again in the second line, $\widetilde{\ex}_{\theta} [\partial_{\theta} \ln p_{\theta}^{\tau}] = \widetilde{\ex}_0 [ \partial_{\theta} p_{\theta}^{\tau}] =  \partial_{\theta} \widetilde{\ex}_{\theta} [1]$ and the assumption (\ref{eq:indep}) in the third line, and the Cauchy-Schwarz inequality in the fourth line. By using a similar argument, we find that
\begin{align}
 \widetilde{\ex}_{\theta} [ (\partial_{\theta} \ln p_{\theta}^{\tau} )^2] = \widetilde{\ex}_{\theta} [ - \partial_{\theta}^2 \ln p_{\theta}^{\tau} ].
\end{align}
Hence, we have the information inequality at the stopping time:
\begin{align} \label{eq:cr2}
 \frac{ (\partial_{\theta} \widetilde{\ex}_{\theta} [ A_{\tau} ])^2}{ \widetilde{\var}_{\theta} [ A_{\tau} ]} \leq \widetilde{I}_{\tau}(\theta),
\end{align}
where we have defined the modified Fisher information $\widetilde{I}_{\tau}(\theta) \coloneqq \widetilde{\ex}_{\theta} [ - \partial_{\theta}^2 \ln p_{\theta}^{\tau} ]$. If the stopping time $\tau$ is bounded with probability one for any parameter $\theta$, the above inequality simplifies to form (\ref{eq:cr}).

\section{Derivation of KUR (\ref{eq:result}) and (\ref{eq:result2})}
\label{ap:2}

Let $(X_t)_{t \in [0, \infty)}$ be a time-homogeneous and continuous-time Markov chain on a directed multigraph $G = (S, E)$. Here $S$ denotes a discrete state space and $E$ the set of directed edges between two states in $S$. The transition rate associated with the edge $e$ directed from $x$ to $y$ is given by $k_e (x,y)$. The initial data $X_0$ is chosen as $x_0 \in S$. We define the modified processes with transition rates $k_{e,\theta} \coloneqq (1+\theta) k_e(x,y)$, where $\theta$ is the real parameter. The escape rates for the modified process are given by $\lambda_{\theta}(x) = (1+\theta) \lambda (x)$. This modification of the rate corresponds to a rescaling of time in the stochastic process. We use $P_{\theta}$ to denote the path probability distribution for the modified process. The distribution $P_0 \eqqcolon P$ corresponds to the original stochastic process.

For a time-integrated current $J_t$ of the form (\ref{eq:current}), we consider the first passage time $\tau \coloneqq \inf \{ t \geq 0 : J_t > J_{\mathrm{th}} \}$ at which the accumulated current $J_t$ first reaches a certain threshold $J_{\mathrm{th}}$. The observable $J_t$ has the property $J [ (Y_s \coloneqq X_{cs})_{s \in [0, c^{-1} t]} ] = J_t$ for any $c > 0$. This implies that
\begin{align}
 \tau [ (X_{cs})_{s \in [ 0, \infty)}] = c^{-1} \tau [ (X_s)_{s \in [0, \infty)]}].
\end{align}
This means that the first passage time for the time-integrated current scales by factor $c^{-1}$ because of the rescaling of time for the stochastic trajectories $X_s \mapsto X_{cs}$. Therefore, the cumulative distribution function of $\tau$ with respect to $P_{\theta}$ and $P_0$ are related through
\begin{align*}
 P_{\theta, x_0} ( \tau \leq t ) &= \sum_{n = 1}^{\infty} P_{\theta, x_0} (\tau \leq t \land N_t = n)
 \notag \\
 &= \sum_{n=1}^{\infty} \sum_{(x_1, e_1: \dots : x_n, e_n) } \int_{0 < t_1 < \dots < t_n < t} dt_1 \dots dt_n \ 1_{\{\tau \leq t \}} [ (x_s)_{s \in [0, t]}]
 \notag \\
 & \ \ \  \times e^{- \lambda_{\theta} (x_0) \cdot  t_1} k_{e_1,\theta} (x_0, x_1) e^{- \lambda_{\theta} (x_1) \cdot (t_2 - t_1)} \dots k_{e_n,\theta}(x_{n-1}, x_n) e^{-\lambda_{\theta}(x_n) \cdot (t - t_n)}
 \notag \\
 &= \sum_{n=1}^{\infty} \sum_{(x_1, e_1: \dots : x_n, e_n)} \int_{0 < t_1 < \dots < t_n < t} dt_1 \dots dt_n \ 1_{\{\tau \leq t \}} [ (x_s)_{s \in [0, t]}]
 \notag \\
 & \ \ \ \times (1+\theta)^n e^{- \lambda (x_0) \cdot  (1+\theta) t_1} k_{e_1} (x_0, x_1) e^{- \lambda (x_1) \cdot (1+\theta) (t_2 - t_1)} \dots k_{e_n}(x_{n-1}, x_n) e^{-\lambda (x_n) \cdot (1+\theta) (t - t_n)}
 \notag \\
 &= \sum_{n=1}^{\infty} \sum_{(x_1, e_1: \dots : x_n, e_n)} \int_{0 < t_1^{\prime} < \dots < t_n^{\prime} < (1+\theta) t} dt_1^{\prime} \dots dt_n^{\prime} \ 1_{\{\tau \leq (1+\theta) t \}} [ (x_{(1+\theta)^{-1}s})_{s \in [0, (1+\theta)t]}]
 \notag \\
  & \ \ \ \times e^{- \lambda (x_0) \cdot t_1^{\prime}} k_{e_1} (x_0, x_1) e^{- \lambda (x_1) \cdot (t_2^{\prime} - t_1^{\prime})} \dots k_{e_n}(x_{n-1}, x_n) e^{-\lambda (x_n) \cdot ((1+\theta)t - t_n^{\prime})}
  \notag \\
  &= P_{x_0} (\tau \leq (1+\theta) t).
\end{align*}
This relation implies that the unnormalized probability density function $f_{\theta}(t)$ of $\tau$ satisfies $f_{\theta} (t) = (1+\theta) f_0 ((1+\theta)t)$. Hence, we have that $\widetilde{\ex}_{\theta, x_0} [ \tau ] = (1+\theta)^{-1} \widetilde{\ex}_{x_0} [\tau]$ and $\partial_{\theta} \widetilde{\ex}_{\theta, x_0} [ \tau ] = - (1+\theta)^{-2} \widetilde{\ex}_{x_0}[\tau]$.

Moreover, the logarithm of the likelihood ratio function $p_{\theta}^t = dP_{\theta}^t / dP_0^t$ is given by
\begin{align}
 \ln p_{\theta}^t &= \sum_{i=1}^{N_t} \ln \frac{k_{e_i, \theta}(x_{i-1},x_i)}{k_{e_i}(x_{i-1},x_i)} - \int_0^t [ \lambda_{\theta} (X_s) - \lambda (X_s) ] ds
 \notag \\
 &= N_t \ln (1+\theta) - \theta \int_0^t \lambda (X_s) ds
\end{align}
Therefore, we obtain that $- \partial_{\theta}^2 \ln p_{\theta}^t = (1+\theta)^{-2} N_t$ and $\widetilde{I}_{\tau}(\theta) = (1+\theta)^{-2} \widetilde{\ex}_{\theta} [ N_{\tau} ]$.

The rescaling of time with finite factor $(1+\theta)$ does not change the probability that the first passage time diverges, i.e., $\widetilde{\ex}_{\theta}[1]$ is independent of $\theta$. Therefore, we can apply the information inequality (\ref{eq:cr2}) to this modified processes. By taking $\theta = 0$ in (\ref{eq:cr2}), we have the modified KUR (\ref{eq:result2}). If $P (\tau < \infty) = 1$, the KUR (\ref{eq:result}) is recovered.

\section{Mean dynamical activity in random walk}
\label{ap:3}

The random walker $X_t$ is described in the form,
\begin{align}
 X_t = \sum_{i=1}^{N_t} Z_i.
\end{align}
Here $N_t$ is the number of jumps over $[0,t]$ and $(Z_1, Z_2, \dots)$ are random variables independently and identically distributed according to $P(Z_i = \pm 1) = k_{\pm} / (k_{+} + k_{-})$. Let us consider the first passage time $\tau_x = \inf \{ t \geq 0 : X_t = x \}$. Because condition $X_t = x$ is equivalent to $\sum_{i=1}^{N_t} Z_i = x$, $N_{\tau_x}$ is a first passage time at which the discrete-time stochastic process $(Z_1, Z_2, \dots)$ first reaches the threshold $x$. If $x > 0$, the expectation value of $N_{\tau_x}$ is finite. By applying the Wald identity \cite[Theorem 2.4.4]{GMS}, we obtain $\ex [X_{\tau_x}] = \ex [ N_{\tau_x}] \cdot \ex [Z_i]$. Hence, 
\begin{align}
 \ex [ N_{\tau_x} ] = \frac{\ex [X_{\tau_x}]}{\ex [Z_i]} = \frac{x}{\tanh(\epsilon /2)}.
\end{align}

\end{widetext}

\end{document}